\documentclass[
prl, twocolumn,
superscriptaddress,
nofootinbib,
 amsmath,amssymb,
 aps,
floatfix,
]{revtex4-1}

\usepackage{graphicx}
\usepackage{textcomp} 
\usepackage{gensymb} 
\usepackage{dcolumn}
\usepackage{bm}
\usepackage{amsmath, amssymb}
\usepackage{dsfont}
\usepackage{float}
\usepackage{afterpage}
\usepackage{xcolor}
\usepackage[normalem]{ulem}

\newcommand{\ket}[1]{\vert #1 \rangle}
\newcommand{\bra}[1]{\langle #1 \vert}
\newcommand{\braket}[2]{\langle #1 \vert #2 \rangle}
\newcommand{\abs}[1]{\left|\left| #1 \right| \right|}

\begin{document}


\title{Quantum Correlations in the Kerr Ising Model}

\author{M. J. Kewming} 
\email{michael.kewming@gmail.com}
\affiliation{Centre for Engineered Quantum Systems, School of Mathematics and Physics, University of Queensland, QLD 4072 Australia}
\author{S. Shrapnel}
\affiliation{Centre for Engineered Quantum Systems, School of Mathematics and Physics, University of Queensland, QLD 4072 Australia}
\author{G. J. Milburn}
\affiliation{Centre for Engineered Quantum Systems, School of Mathematics and Physics, University of Queensland, QLD 4072 Australia}

\date{\today}
\begin{abstract}
In this article we present a full description of the quantum Kerr Ising model---a linear optical network of parametrically pumped Kerr non-linearities. We consider the non-dissipative Kerr Ising model and, using variational techniques, show that the energy spectrum is primarily determined by the adjacency matrix in the Ising model and exhibits highly non-classical cat like eigenstates. We then introduce dissipation to give a quantum mechanical treatment of the measurement process based on homodyne detection via the conditional stochastic Schrodinger equation. Finally, we identify a quantum advantage in comparison to the classical analogue for the example of two anti-ferromagnetic cavities. 
\end{abstract}

\maketitle
\section*{Introduction}
In an increasingly connected and dynamic society, the demand to solve complex problems requiring optimal configurations of large systems is growing. Many of these optimisation problems fall into the NP-hard or NP-complete complexity class that are practically impossible to solve on a classical digital computer. These types of problems can be mapped onto the Ising model where the ground state yields the optimal solution. Attempts to find the ground state has lead to the development of many classical and quantum approaches including adiabatic quantum computing \cite{farhi_quantum_2001} and quantum annealing \cite{kadowaki_quantum_1998,brooke_quantum_1999}. However, a significant hurdle facing these architectures is the connectivity of individual physical qubits, an essential requirement for solving large optimisation problems.

A new approach dubbed the Coherent Ising Machine (CIM) overcomes this issue by implementing an optical analog of the Ising model \cite{wang_coherent_2013,takata_quantum_2015, mcmahon_fully_2016, yamamoto_coherent_2017}. Recent results have compared the performance of the CIM against semi-definite programs and simulated annealing \cite{haribara_coherent_2016} as well as neural network architectures \cite{haribara_performance_2017} and quantum annealers \cite{hamerly_experimental_2019}. Further theoretical results have shown comparable results can be obtained when modelled using Gaussian optics \cite{clements_gaussian_2017}. In the CIM, a series of degenerate parametric oscillators (DOPO's) form a coupled network of potential spins. At sufficiently high pumping, the DOPO experiences a pitchfork bifurcation into two steady state solutions. These two solutions are coherent states which are $\pi$ out of phase with one another and play the role of a `spin' in the Ising model. When coupled, each DOPO in the network bifurcates in accordance with an Ising Hamiltonian. The lowest energy configuration of coherent states in the cavities therefore corresponds to the ground state. The network of DOPO's become non separable during bifurcation \cite{takata_quantum_2015, maruo_truncated_2016} which may lead to a quantum advantage \cite{inagaki_coherent_2016}, although it is not yet clearly identified. Recently, several impressive classical experimental realisations of the CIM have been published demonstrating up to $2000$ connected spins \cite{mcmahon_fully_2016,inagaki_coherent_2016, hamerly_experimental_2019}. Furthermore, off-the-shelf electronics have been used to build robust opto-electronic CIMs \cite{bohm_understanding_2018}. 

Here we will consider a Kerr Ising Model (KIM) based on a network of parametrically pumped cavities containing a Kerr non-linearity.  
This interaction has been studied extensively with seminal results showing the onset of chaos \cite{milburn_quantum_1991}, single and multi-photon blockade \cite{leonski_possibility_1994, imamoglu_strongly_1997, miranowicz_two-photon_2013, miranowicz_state-dependent_2014}, as well as qubit construction \cite{mabuchi_qubit_2012}, quantum gates \cite{pachos_optical_2000, knill_scheme_2001,nysteen_limitations_2017} and computation \cite{goto_universal_2016,combes_two-photon_2018}. Furthermore, experimental realisations of parametrically driven of Kerr non-linearities and their subsequent quantum behaviour have been discussed \cite{puri_engineering_2017} and realised in superconducting circuits \cite{wang_quantum_2019}. There is now a small but growing literature on driven networks of Kerr non-linearities in cascaded systems \cite{stannigel_driven-dissipative_2012}, under adiabatic evolution \cite{goto_bifurcation-based_2016, goto_universal_2016, goto_quantum_2019} and dissipative single photon loss \cite{puri_quantum_2017, nigg_robust_2017}. Notably, the results presented in \cite{goto_bifurcation-based_2016, goto_quantum_2019} describe the implementation of an Ising machine constructed from a network of Kerr non-linearities but do not consider the nature or effects of any quantum correlations; an analysis we present here.

We start by presenting a comprehensive theoretical description of the KIM and show that the network exhibits non-classical correlations at the corresponding classical bifurcation --- indicating a quantum phase transition \cite{hines_quantum_2005}. Beyond the bifurcation the ground state corresponds to an entangled state of minimum energy spin configurations of the Ising model. We then introduce low temperature dissipation using the quantum optics master equation and show that the steady state corresponds to a probabilistic mixture of these highly entangled ground states. This enables us to describe the conditional dynamics corresponding to continuous homodyne detection of the output fields showing the stochastic approach to the desired ground state. 
Lastly, we compare the fully quantum model against a classical analogue at finite temperatures for two coupled cavities and clearly identify the advantage arising from quantum correlations. 

\begin{figure}
    \centering
    \includegraphics[width=\columnwidth]{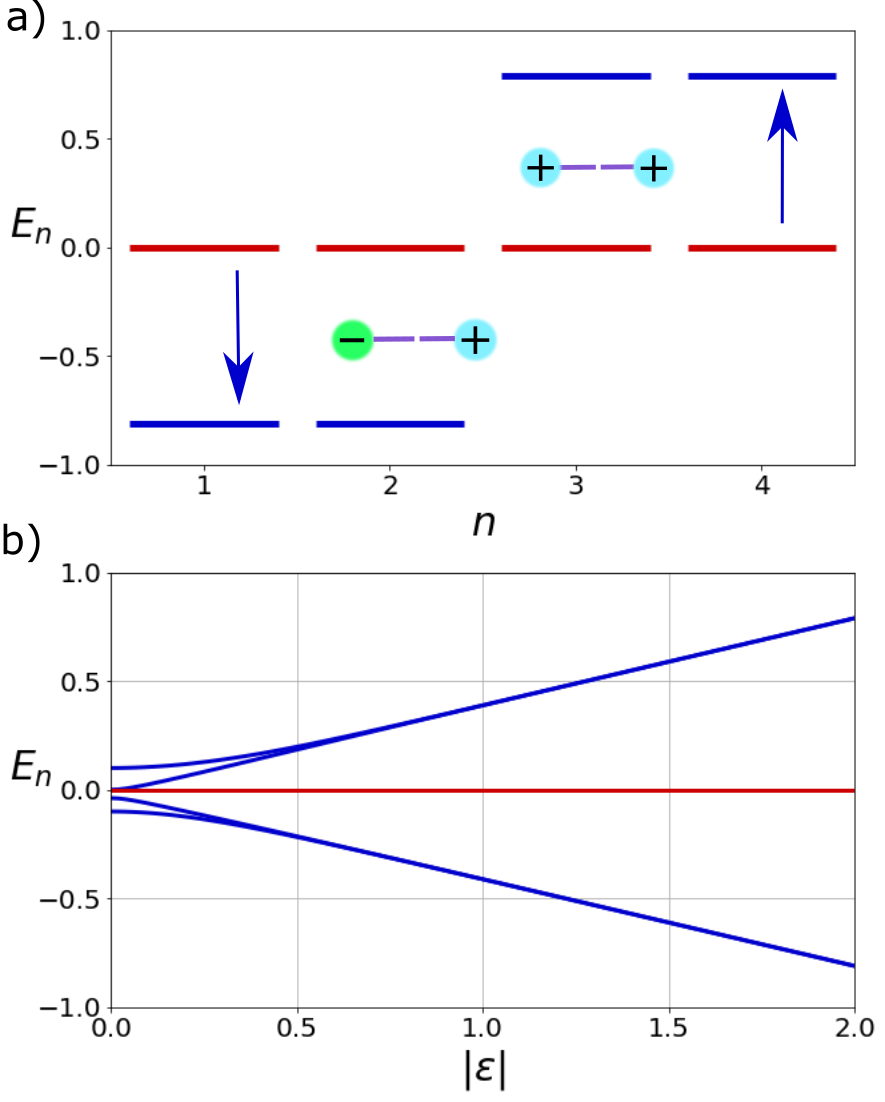}
    \caption{a) Energy of the $n$th eigenstate $E_{n}$ of 2 spins---2 coupled cavities with no detuning $\Delta=0$. Red lines correspond to the uncoupled network $\eta=0$ and blue correspond to $\eta = 0.05$ with driving strength $\vert\epsilon\vert = 2$. When the network is uncoupled, there are $4$ degenerate ground states corresponding to each unique configuration of spins. However, when a small coupling is introduced, the degeneracy is broken leading to $2$ new ground states and $2$ excited states in accordance with the Ising model. b) For weak coupling, the energy levels change linearly according to the perturbed energy levels Eq.\,(\ref{eq:perturbation}). 
    \label{fig:energy_levels}}
\end{figure}

\section{The Kerr Ising Model}
In our analysis, we will start by assuming that the pump mode has been adiabatically eliminated. In the interaction picture, each cavity in the network is described by the Hamiltonian
\begin{equation}
\label{eq:ppk}
    H_{0} = \hbar\chi \left( a^{\dagger 2} - \frac{\epsilon^{*}}{\chi}\right)\left(a^{2} - \frac{\epsilon}{\chi}\right) + \hbar\Delta a^{\dagger}a
\end{equation}
where $\chi$ is proportional to the third-order nonlinear susceptibility, $\Delta=\omega_a-\omega_p/2$ is the detuning of the pump frequency, $\omega_p$,  from cavity resonance frequency, $\omega_a$, and $\epsilon$ is pump field. 

Due to the parity symmetry of Eq.(\ref{eq:ppk}), on resonance the ground state is a doubly degenerate subspace with energy ${E_{0} = 0}$ spanned by two coherent states $\ket{\alpha_{i}} {= \ket{\pm \alpha_{0}}}$ where ${\alpha_{0} = \sqrt{\epsilon/\chi}}$. The introduction of a \emph{small} detuning $\Delta$ creates a linear shift in the energy levels $E_{0} {= \hbar \Delta \vert \alpha_{0} \vert^{2}}$. Likewise, for a network of $N$ \emph{uncoupled} Kerr-cavities the ground state is $2N$-fold degenerate spanned by product state $\ket{\vec{\alpha}_{m}} = \ket{\pm \alpha_{0}, \pm \alpha_{0}, ...}$ with a perturbed energy of $E_{0}^{'} = \hbar N \Delta \vert \alpha_{0} \vert^{2}$. 

If we now introduce a linear coherent coupling scheme in the form of a network of connected beam-splitters and phase shifters ---we obtain the Hamiltonian \cite{Milburn:07}
\begin{equation}
    H_{\mathrm{ising}} = \sum_{i}^{N} H_{0}^{(i)} + \hbar \eta \,\vec{a}^{\dagger}.\hat{S}.\vec{a}
\end{equation}
where $\eta$ is fixed coupling strength for each cavity, $S$ is the adjacency matrix between cavities and $\vec{a} = (a_{1}, a_{2},...,a_{N})$ is a vector representation of annihilation operators for each cavity. 
On closer inspection, $H$ bears considerable resemblance to the classical Ising model of $N$ coupled spins $\sigma_{i} = \pm 1$ in a uniform magnetic field $B$
\begin{align}
    H_{I} &= H_{B} -J \sum_{\langle i,j \rangle} \sigma_{i}\sigma_{j}\, \nonumber\\
    \label{eq:Ising}
    & = H_{B} - J \vec{\sigma}^{T}.\hat{S}.\vec{\sigma}\,,
\end{align}
where $H_{B} = - \mu B \sum_{i}\sigma_{i}$ is the magnetic dipole Hamiltonian, $J$ is the uniform coupling between spins and $\hat{S}$ is again, the adjacency matrix between spins. In the KIM, the parametric driving of each Kerr non-linearity creates a subspace of pseudo-spins coupled through a linear optical network. This model mimics the magnetic field driving and the magnetic coupling between spins in the traditional Ising model. As a result, the energy spectrum in the KIM will reflect similar characteristics to the Ising model in Eq.(\ref{eq:Ising}). 

Using degenerate perturbation theory for a weak coupling $\eta$ and large driving $\epsilon> 0$, the $n$th shifted energy level of the network is determined by
\begin{equation}
\label{eq:perturbation}
    E_{n}^{'} = \hbar N \Delta \frac{\vert \epsilon \vert}{\chi} +  \hbar \eta \vec{\alpha}_{n}^{T}.\hat{S}.\vec{\alpha}_{n}
\end{equation}
where $\vec{\alpha}_{n} = \vert \epsilon \vert /\chi \vec{\sigma}_{n}$ corresponds to the configuration vector of cavity coherent states for the $n$th eigenstate $\ket{\vec{\alpha}_{n}}$ \cite{goto_bifurcation-based_2016}. The eigenstates of the new spectrum can be found by solving the perturbed eigenvector equation
\begin{equation}
    \eta \,\vec{a}^\dagger.\hat{S}.\vec{a} \ket{\vec{\alpha}_{n}^{'}} = \eta\, \vec{\alpha}_{n}^{\dagger}.\hat{S}.\vec{\alpha}_{n} \ket{\vec{\alpha}_{n}^{'}}\,,
\end{equation}
where $\ket{\vec{\alpha}^{'}} = \sum_{i}^{2N}c_{ni}\ket{\vec{\alpha}_{i}}$ is written as a linear combination of the uncoupled coherent eigenstates $\ket{\vec{\alpha}_{i}}$. Multiplying through by $\bra{\vec{\alpha}_{k}}$ and using the annihilation operator eigenvalue equation $\vec{a} \ket{\vec{\alpha}} = \vec{\alpha} \ket{\vec{\alpha}}$ we obtain the matrix equation
\begin{equation}
\label{eq:eigen}
      c_{nk} \frac{\vert \epsilon \vert}{\chi}\eta \left( \vec{\sigma}_{k}^{\dagger}.\hat{S}.\vec{\sigma}_{k}\,  - \vec{\sigma}_{n}^{\dagger}.\hat{S}.\vec{\sigma}_{n} \right)  = 0\,.
\end{equation}
As $c_{ik} \neq 0$, this relationship is satisfied so long as $ \vec{\alpha}_{k}^{\dagger}.\hat{S}.\vec{\alpha}_{k} = \vec{\alpha}_{n}^{\dagger}.\hat{S}.\vec{\alpha}_{n} $.
This ensures any linear superposition of spin configurations $\vec{\alpha_{n}}$ which have the same energy will form a degenerate subspace. Given the parity symmetry in the Hamiltonian, the subspace will be spanned by the orthogonal cat-like superpositions of the Ising solutions.

To illustrate this, we consider the simplest 2-spin Ising model for zero detuning $\Delta=0$ shown in Fig.(\ref{fig:energy_levels}). In the absence of coupling $\eta=0$, the ground state is $4$-fold degenerate spanned by every unique configuration of spins $\ket{\alpha_{i}}$.

Now introducing a weak coupling, the degeneracy is broken into a $2$-fold degenerate ground state spanned by the anti-ferromagnetic states ${\ket{\lambda_{-}} \propto \ket{\alpha_{0},-\alpha_{0}} \pm \ket{-\alpha_{0},\alpha_{0}}}$ and a $2$-fold degenerate excited state spanned by the ferromagnetic states ${\ket{\lambda_{+}} \propto \ket{\alpha_{0},\alpha_{0}} \pm \ket{-\alpha_{0},-\alpha_{0}}}$.
The new ground state in the high driving limit has an energy $E_{g}^{'} = -2\hbar \eta \vert \epsilon \vert/\chi$ whereas the excited state has energy $E_{e}^{'} = 2 \hbar \eta \vert \epsilon \vert/\chi$. 
 
A notable feature is the limit of low driving plotted in Fig.(\ref{fig:energy_levels}b) where the system is not degenerate. In this limit, the coupling $\eta$ perturbation is strong compared to the driving $\epsilon$ and breaks the degeneracy.  

The evolution of the KIM can be described by the Von-Neumann time development equation
\begin{equation}
    \frac{d\rho(t)}{dt} = -\frac{i}{\hbar} \left[H_{\mathrm{ising}},\rho\right]\,,
\end{equation}
This may be converted to a Fokker-Planck equation for the positive P-representation of $\rho$ using an expansion of the density operator over the off-diagonal projectors $|\vec{\mu}\rangle\langle \vec{\nu}|$,
\begin{align}
\label{eq:fokker-planck}
    \frac{\partial P(\vec{\mu}, \vec{\nu})}{\partial t} &= -\partial_{\vec{\mu}}\left[\Omega(\vec{\mu},\vec{\nu}) P(\vec{\mu}, \vec{\nu})\right] + \partial_{\vec{\nu}}\left[\Omega(\vec{\nu},\vec{\mu}) P(\vec{\mu}, \vec{\nu})\right] \nonumber\\
    & + \partial_{\vec{\mu}}^{2} \left[\Phi(\vec{\nu},\vec{\mu}) P(\vec{\mu}, \vec{\nu})\right]-  \partial_{\vec{\nu}}^{2} \left[\Phi(\vec{\nu},\vec{\mu})P(\vec{\mu}, \vec{\nu})\right]\,,
\end{align}
where the diffusion term is $\Phi(\vec{\mu},\vec{\nu}) = -i(\chi \vec{\mu}^{2} + \epsilon)$ and the drift is $\Omega(\vec{\mu},\vec{\nu}) = -i(2\chi \vec{\mu}^{2} \vec{\nu} -2\epsilon \vec{\nu} + \Delta\vec{\mu} + \eta \hat{S}.\vec{\mu})$.

Replacing $\vec{\mu} = \vec{\alpha}$ and $\vec{\nu} = \vec{\alpha}^{*}$ and ignoring the quantum diffusion terms yields the semi-classical equations of motion
\begin{equation}
\label{eq:classical}
    \frac{\partial}{\partial t} 
\left(\begin{array}{c}
        \vec{\alpha} \\
        \vec{\alpha}^{*}
    \end{array}
\right)
=
i \hat{A} 
\left(\begin{array}{c}
        \vec{\alpha} \\
        \vec{\alpha}^{*}
    \end{array}
\right)
\end{equation}
where 
\begin{equation}
    \hat{A} = 2 \epsilon \mathbf{I}_{N} \otimes \sigma_{x} - \left(\Delta\, \mathbf{I}_{N} + 2 \chi \vert \alpha \vert^{2} \mathbf{I}_{N} +  \eta  \hat{S}\right) \otimes \sigma_{z}
\end{equation}
where $\hat{\sigma}_{x,y,z}$ are the Pauli spin matrices. The structure of $\hat{A}$ is analogous to the spin model coupled along $\hat{\sigma}_{z}$ and driven along $\hat{\sigma}_{x}$  where the Hilbert space of the cavities constructs the pseudo-spin structure and interactions.
The bistability of the system can then be computed when a sign change occurs in the trace $\mathrm{tr}(\hat{A})$ or the determinant $\det(\hat{A})$ with the variation of the driving $\epsilon$ \cite{drummond_quantum_1980}.

\section{Entanglement in the KIM}

\begin{figure*}
    \centering
    \includegraphics[width=2\columnwidth]{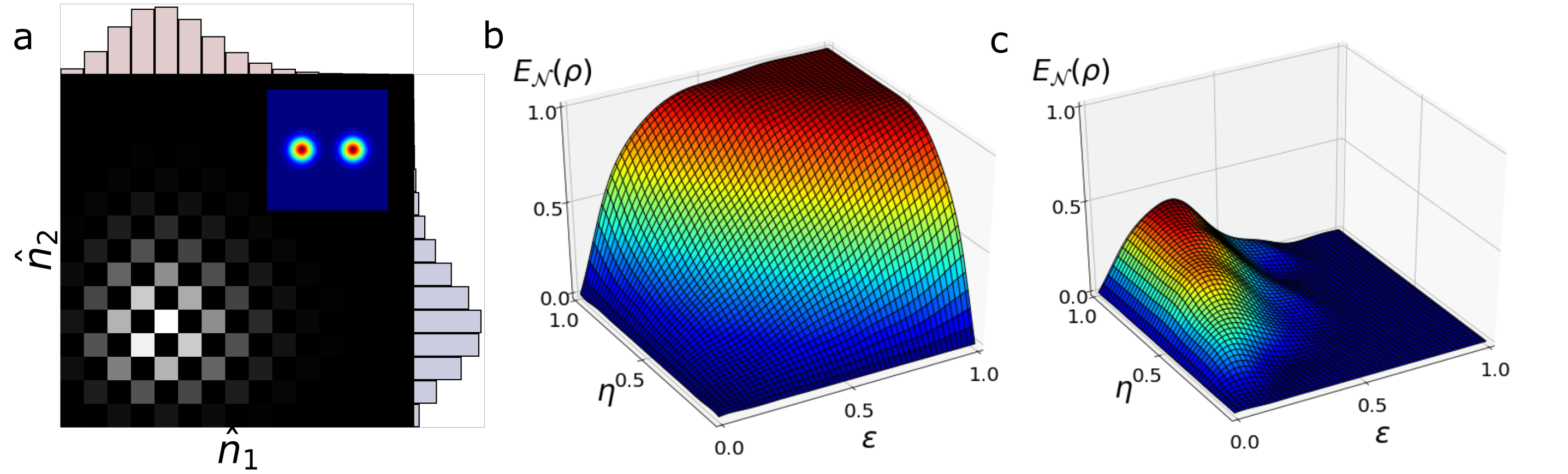}
    \caption{a) Joint photon number distribution with for two entangled cavities with parameters $\chi=0.5$, $\eta = 0.1, \epsilon = 1.5, \Delta=1$: detuning pushes the bifurcation way from $\epsilon>0$. b) The LN of an individual cavity's ground state in the KIM. As the driving increases the ground state becomes a highly entangled cat like state satisfying Eq.(\ref{eq:eigen}). c) The LN of the steady state if the cavities are damped into a zero temperature heath bath ( $\bar{n}=0$). The correlations between the cavities become non-classical as the system undergoes bifurcation but rapidly decay to classical correlations (zero entanglement)  in the large driving limit.}
    \label{fig:entanglement}
\end{figure*}

\subsection{Ground state entanglement}

The KIM has an energy spectrum analogous to the Ising model of a spin network immersed in a constant magnetic field. Earlier work on the quantum Ising model has shown the ground state undergoing a quantum phase transition, through which the degenerate ground states become highly entangled \cite{hines_quantum_2005}. Likewise, previous dissipative CIM models have also shown that the entanglement and quantum discord occur in the pseudo-spin network \cite{takata_quantum_2015, maruo_truncated_2016}.

There are many different tests and criterion for quantum correlations and entanglement. One such measure of non-separability of Gaussian states was introduced by Duan \emph{et al} \cite{duan_inseparability_2000} and used in the dissiptive CIM models \cite{takata_quantum_2015,maruo_truncated_2016}. This criterion is a suitable choice in dissipative models if the steady state solution is Gaussian.
The eigenstates of the KIM are highly non-Gaussian cat like states seen in the joint photon number distribution of $\braket{\hat{n}_{1},\hat{n}_{2}}{\lambda_{0}}$ plotted in Fig.(\ref{fig:entanglement})a which exhibits significant interference suggesting the presence of strong non-classicality. Given the non-Gaussian nature of the eigenstates, using the Duan inequality is not a necessary condition of inseparability \cite{duan_inseparability_2000}. A more robust metric to compare between Hamiltonian and dissapative models is the \emph{logarithmic negativity} (LN) defined as 
\begin{equation}
    E_{\mathcal{N}}(\rho) = \log_{2} \abs{\rho^{T_{A}}}_{1}
\end{equation}
where $\abs{\rho^{T_{A}}}_{1} = 2\mathcal{N} +1$ is the trace norm defined in terms of the \emph{negativity} $\mathcal{N}$ which is the absolute sum of all negative eigenvalues of $\rho^{T_{A}}$ \cite{vidal_computable_2002}. In the limit of maximally entangled pure states, the LN reaches equality with the Von-Neumann entropy metric of entanglement. 

Plotting the LN in Fig.(\ref{fig:entanglement})b for a large non-zero detuning $\Delta=1$ we see that the non-resonant pumping of the cavity pushes the bifurcation away from $\epsilon=0$ and plays an analogous role to the damping considered in the dissapative models---which we will consider in the next section---but retains the unitarity of the system. The ground state smoothly transitions to a highly entangled state and remains so for increased driving $\epsilon$ and coupling $\eta$. 

\subsection{Steady state entanglement}

In an experimental setting, one needs to measure the values of each pseudo-spin in the  model and thus the closed Hamiltonian system must be made open in some way.  This could be an impulsive coupling to the external apparatus after some period of unitary evolution. In Ref.\cite{goto_bifurcation-based_2016} the solution to the model was obtained under adiabatic evolution followed by a final read out. In quantum optics however a more conventional approach would be to open the system by letting a small amount of light leak out of the cavity and subject that to continuous measurement. Once the cavity is opened in this way the system becomes a damped non linear system and the elliptic fixed points of the Hamiltonian now become stable fixed points of a dissipative dynamical system. As for what to measure, we note that the key feature of the KIM solutions we seek are in the phase of the intracavity field so we will need to consider a phase-dependent measurement scheme such as homodyne detection. Given a measurement record we can now ask for the conditional dynamics of the intracavity field given a given stochastic homodyne current record. 

At a non-zero temperature---where the heat bath is treated as white-noise on the input field---and for unit quantum efficiency homodyne detection, the conditional state obeys the conditional Stochastic Schrodinger equation (SSE) \cite{WisemanH.M.HowardM.2010Qmac}
\begin{eqnarray}
\label{eq:stochastic}
    d\rho(t) & = & -\frac{i}{\hbar} \left[H_{\mathrm{ising}}, \rho(t)\right]dt + \sum_{i}^{N}\gamma(\bar{n} + 1) \mathcal{D}[a_{i}]\rho dt \\
    & & + \gamma \bar{n} \mathcal{D}[a_{i}^{\dagger}]\rho dt + \sqrt{\gamma} \mathcal{H}[\gamma(\bar{n}+1)a_{i} - \gamma \bar{n} a_{i}^{\dagger}]dW_{i}(t)\nonumber \,,
\end{eqnarray}
where all photons emitted from the cavity at rate $\gamma$ are detected and $dW_{i}(t)$ is the classical Weiner process. The mean-photon number due to thermal excitation is determined by the Bose-Einstein statistics $\bar{n} = (\exp\left[{\hbar \omega/k_{b}T}\right]-1)^{-1}$ where $T$ is the temperature and $\omega$ is the cavity resonance frequency (assumed to be identical). Here, $\mathcal{D}[a_{i}]\rho$ is the single photon loss channel and $\mathcal{H}[a]\rho =  a \rho + \rho a_{i}^{\dagger} - \mathrm{tr}\left(a_{i} \rho + \rho a_{i}^{\dagger}\right)\rho$ is the conditional stochastic term of the $i$th cavity. If we only seek the unconditional dissipative evolution, we can average over the stochastic term in Eq.(\ref{eq:stochastic}) which then vanishes giving a master equation for the state of the damped cavity field. This master equation may now be  converted to an equivalent Fokker-Planck equation as Eq.(\ref{eq:fokker-planck}) with the addition of a decay term introduced into the drift $\Omega(\vec{\mu},\vec{\nu}) \rightarrow \Omega(\vec{\mu},\vec{\nu}) - \gamma \vec{\mu}/2$ and an additional noise term $\gamma \bar{n}\,\partial_{\vec{\nu} \vec{\mu}}^{2}P(\vec{\mu},\vec{\nu})$. Instead of asking about the structure of the ground state our attention now turns to the steady state solution to the master equation or equivalently to the Fokker Planck equation. 

The LN of the steady state solution of two cavities as a function of pumping $\epsilon$ and coupling $\eta$ is depicted in Fig.\, (\ref{fig:entanglement})c at zero temperature $\bar{n}=0$. The entanglement is reduced due to the introduction of cavity dissipation but is non zero at the bifurcation. The two cavities become non-separable around parameter values corresponding to the classical bifurcation in the semi-classical dissipative model indicating a dissipative quantum phase transition \cite{hines_quantum_2005}. The entanglement then decreases to zero in the limit of large driving. A similar peak in non-separability was also observed in the CIM \cite{takata_quantum_2015, maruo_truncated_2016}. 

In the case of the conditional dynamics based on homodyne detection, the cavities stochastically evolve into one particular spin configuration $\ket{\pm\alpha_{0},\mp\alpha_{0}}$.
One can show the steady state solutions correspond to the different possible spin configuration sin the Ising model determined by the adjacency matrix $\hat{S}$. This is achieved by finding the roots of Eq.\ref{eq:classical} when $\partial \vec{\alpha}/\partial t = 0$ \cite{yamamoto_coherent_2017, nigg_robust_2017}. 
As we can regard the unconditional steady state as arising from ensemble averaging over the measurement records and thus the conditional states, we conclude in the large driving limit, the steady state can be regarded as a mixture of the possible spin-state configurations that satisfy Eq.(\ref{eq:eigen}), slightly shifted due to the damping.

\section{Comparison between Quantum and Classical models}

In the previous section we outlined the quantum dynamics of the 2 spin anti-ferromagnetic KIM subject to weak damping so as to allow constant homodyne detection through the conditional SSE Eq.\,(\ref{eq:stochastic}). 
By modelling the conditional SSE we provide a numerical experiment of results that would likely be observed in a physical realisation of the KIM.
We seek to understand more carefully what advantage may arise at the quantum limit. This can be achieved by making a comparison between the semi-classical model and the full quantum model at finite temperature as the temperature is lowered. Here we will compare the conditional field amplitude averages computed by solving the conditional SSE model with stochastic solutions to  the SDE described by Eq.\,(\ref{eq:classical}) with the addition of thermal noise 
\begin{equation}
\label{eq:classical_sde}
    \frac{\partial}{\partial t} 
\left(\begin{array}{c}
        \vec{\alpha} \\
        \vec{\alpha}^{*}
    \end{array}
\right)
=
i \hat{A} 
\left(\begin{array}{c}
        \vec{\alpha} \\
        \vec{\alpha}^{*}
    \end{array}
\right) 
+ \sqrt{\frac{\gamma \bar{n}}{2}} 
\left(\begin{array}{cc}
        i & 1 \\
        -i & 1
    \end{array}
\right) 
\left(\begin{array}{c}
        \xi_{1}(t) \\
        \xi_{2}(t)
    \end{array}
\right)\,, 
\end{equation}
where $\xi_{1}(t)$ and $\xi_{2}(t)$ are delta correlated stochastic forces.
The measured homodyne currents from either cavity, in the quantum case is ${J_{i}(t) = \langle \hat{X}_{i}(t) \rangle_c + \sqrt{\gamma(2\bar{n}+1)}\xi_{i}(t)}$  where the first term is a quantum average  that is computed from the conditional state. In the classical case the analogous quantity is simply $j_i(t) = \alpha_i(t)+\alpha^*_i(t)$.

\begin{figure*}
    \centering
    \includegraphics[width=\columnwidth]{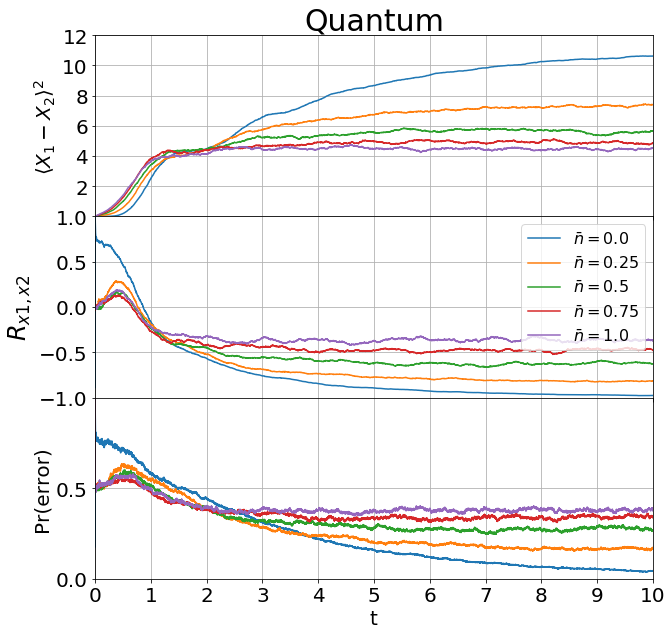}
    \includegraphics[width=\columnwidth]{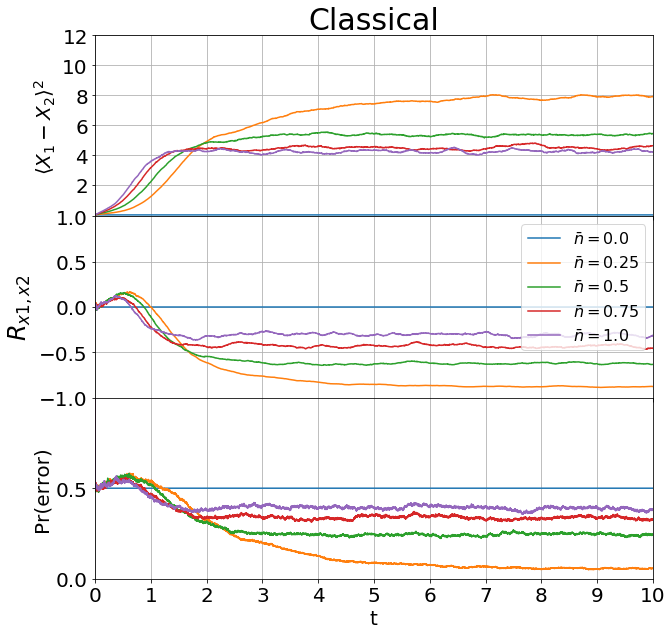}
    \caption{Results from the homodyne currents measured by the quantum mechanical SSE Eq.\,(\ref{eq:stochastic}) (left) and the classical SDE Eq.\,(\ref{eq:classical_sde}) (right). The top figure depicts the mean difference squared in the conditional stochastic mean quadrature amplitudes   (homodyne current minus the noise) as a function of time. In the classical model at zero temperature $\bar{n}=0$, the system remains fixed at the unstable fixed point at the origin. The middle figure is a measurement of the normalised cross correlation function $R_{X_{1},X_{2}}$. At zero the quantum trajectories are strongly positively correlated likely due to the non-classical squeezing in the cavities. The peak in positive correlations occurs when the system bifurcates and is present in both the classical and quantum models at finite temperature. The bottom figure shows the probability of measuring an error $\mathrm{Pr(error)}$ from the homodyne current. The positive correlations lead to a peak in errors but are then rapidly suppressed as the cavities settle into opposing `pseudo-spin' states. The zero temperature limit again exhibits large probability of detecting an error prior to bifurcation due to the non-classical correlations.}
    \label{fig:quantum}
\end{figure*}

There are several quantities determined directly from the homodyne current capable of informing us about the performance of the KIM. Firstly, consider the difference squared of the measured currents of either cavity; ignoring the white noise terms this gives, ${( J_{1}(t) - J_{2}(t) )^{2}=\langle \hat{X}_{1} - \hat{X}_{2} \rangle_c^{2}}$. (Note the square of a conditional average). In the anti-ferromagnetic example, if the two cavities converge on the same signed steady state then the mean difference will be zero. Conversely, if the two cavities settle into the anti-ferroamgnetic solution then this quantity will reach $2\vert \alpha \vert^{2}$ since the cavities bifurcate along $\hat{X}$. Notably, this quantity will also give us a measure of the intracavity dynamics.

Secondly the correlations between the individual homodyne currents ${J}_{1}(t)$ and $J_{2}(t)$ is fundamental to understanding the consequences of the quantum correlations ascertained in Fig.(\ref{fig:entanglement})c. Here we measure the normalised cross-correlation function between the two cavities defined
\begin{equation}
    R_{\hat{X}_{1},\hat{X}_{2}} = \frac{1}{M} \sum \frac{\left( J_{1}^{(m)}(t) - \overline{J_{1}(t) }\right)\left( J_{2}^{(m)}(t) - \overline {J_{2}(t) } \right)}{\sqrt{\sigma_{J_{1}}^{2}\,\sigma_{J_{2}}^{2}}}    
\end{equation}
where $M$ is the number of sampled currents and $\sigma_{J_{i}}^{2}$ is the variance in the current of the $i$th cavity. The correlation function is bounded $-1\leq R_{X_{1},X_{2}} \leq 1$ where equality signifies perfect (anti-)correlation. 

Finally, the `pseudo-spin' of the KIM is determined by the sign of the current $J_{i}(t)$. An error arises when both cavities have the same sign $\mathrm{sign}[J_{1}(t)] = \mathrm{sign}[J_{2}(t)]$ in the anti-ferromagnetic example. Thus we can calculate the probability of obtaining an error as a function of time and measure the success probability of the KIM. 

Here we sample both the quantum Eq.\,(\ref{eq:stochastic}) and the classical Eq.\,(\ref{eq:classical_sde}) trajectories 2000 times---each equivalent to an experimental trial. From this sample we compute the average of the quantities depicted in Fig.\,(\ref{fig:quantum}). Both simulations have identical cavity parameters of $\epsilon=1, \Delta=0, \chi = 0.5, \gamma = 1, \eta = 1$ and are varied over several temperatures from $0\leq\bar{n}\leq1$. 

The results exhibit several interesting features. Firstly, increasing the temperature leads to expected agreement between the classical and quantum models as they approach the steady state.
The system experiences bifurcation approximately where the cross-correlation peaks $R_{X_{1}, X_{2}}$. Before bifurcation, the two cavities exhibit \emph{positive} correlations present in both classical and quantum models. This leads to an increase in detecting an error but rapidly decays once the system has bifurcated. In the limit of zero temperature (blue) the classical model stalls in the absence of thermal fluctuations: the KIM remains stuck at the unstable fixed point centred at the origin. The quantum mechanical model on the other hand still continues to function due to the spontaneous emission of the cavities. In-fact at zero temperature, it is less-likely to lead to an error in the steady state $Pr(\mathrm{error)}$. Also, at zero temperature the quantum model exhibits extremely large positive correlations prior to bifurcation as a consequence of the squeezing. This quantum effect is rapidly suppressed in the presence of small amounts of thermal noise.

Another noticeable feature is the time taken to bifurcate. At all temperatures, the quantum mechanical model appears to peak in correlations at $t\sim 0.5$. The mean difference in $\langle X_{1} - X_{2} \rangle^{2}$ appears to plateau between $1<t<2$ before increasing again for larger times. The classical model on the other hand at low temperature---$\bar{n} = 0.25$---is noticeably slower to begin convergence on the Ising solution but begins to resemble the quantum model for $t>2$. In the classical model Fig. (\ref{fig:quantum}), this suggests a critical slowing in the convergence rate to a stable fixed point at lower temperatures. This tension introduces a trade off between speed to undergo bifurcation and accuracy; lower thermal fluctuations lead to higher accuracy but also result in critical slowing of the dynamics. At low temperatures, this critical slowing could be overcome by thermal annealing. 
In the quantum mechanical model at times below $t<2$, the phase transition occurs sooner. This is likely due to the quantum entanglement between the cavities also known as quantum activation \cite{Dykman2006}. 
For larger times---after bifurcation---the cavities transition over into the classical regime as noted by the agreement between the results at finite temperature \cite{maruo_truncated_2016}. 

This indicates one significant advantage over traditional quantum annealing architectures. At zero temperature---a reasonable assumption in optical systems and superconducting circuits---with an initial state in the vacuum, the system is situated at an unstable stationary fixed point. Classically, a mechanism to induce this symmetry breaking is thermal annealing. Quantum mechanically, the cavities can quantum tunnel and begin converging on the stationary solutions determined by the Ising model \cite{yamamoto_coherent_2017}. Therefore \emph{no} annealing is required in the KIM when finding the lowest energy solution.

\section{Discussion}
In this article we have focused on the two-cavity KIM---both classically and the full quantum description. We have shown that the quantum model exhibits a highly entangled ground state but constantly measuring the cavities under homodyne detection forces the system converge on one particular spin-configuration solution. Consequently this leads to a classically correlated steady-state solution whereby quantum correlations exist only during the process of bifurcation. Regardless, the nature of this quantum phase transition elicits a quantum advantage in the trade-off between speed and accuracy of the model as we have shown comparing the classical model to the full quantum description. 

We have shown conclusively that there exists an advantage to using a quantum mechanical KIM at the zero temperature limit as shown in Fig.\,(\ref{fig:quantum}). The quantum fluctuations due to spontaneous emission allow the KIM to operate in the limit where there are no thermal fluctuations. Furthermore, the non-classical correlations arising due to the quantum mechanical squeezing of the joint cavity state lead to an apparent decrease in the time taken for the system to reach bifurcation, likely resulting from quantum activation \cite{Dykman2006}. However it must be noted that this advantage rapidly decays with increasing temperature. Furthermore, we have considered the simplest two cavity model; further investigation is required to determine whether the advantage still exists at a larger number of cavities. It would also be interesting to consider how this advantage manifests itself with more complex adjacency matrices $\hat{S}$ and determine whether or not it is capable of outperforming other classical heuristic Ising solvers. 

\section{Acknowledgements}
The authors would like to thank Joshua Combes for his very insightful comments and suggestions. GJM thanks Hideo Mabuchi for helpful discussions. This research was supported by the Australian Research Council Centre of Excellence for Engineered Quantum Systems (EQUS, CE170100009).

\bibliography{cims}

\end{document}